\documentclass[conference]{IEEEtran}

\usepackage[utf8]{inputenc}
\usepackage{cite}

\usepackage{epsfig}

\makeatletter
\let\MYcaption\@makecaption
\makeatother

\usepackage[font=footnotesize]{subcaption}

\makeatletter
\let\@makecaption\MYcaption
\makeatother

\usepackage{multirow}

\newcommand{\idft}{\mathop{\mathrm{IDFT}}}

\newcommand{\abs}[1]{\left|{#1}\right|}
\usepackage{amsmath}

\usepackage{todonotes}
\usepackage{siunitx}
\usepackage{booktabs}

\DeclareSIUnit{\sample}{S}

\usepackage{glossaries}
\setacronymstyle{long-short}
\newacronym{bs}{BS}{base station}
\newacronym{cir}{CIR}{channel impulse response}
\newacronym{cdf}{CDF}{cumulative distribution function}
\newacronym{ctf}{CTF}{channel transfer function}
\newacronym{dft}{DFT}{discrete Fourier transform}
\newacronym{dl}{DL}{downlink}
\newacronym{idft}{IDFT}{inverse discrete Fourier transform}
\newacronym{los}{LOS}{line of sight}
\newacronym{lpda}{LPDA}{Log-Periodic Dipole Array}
\newacronym{lte}{LTE}{Long Term Evolution}
\newacronym{mimo}{MIMO}{multiple input multiple output}
\newacronym{miso}{MISO}{multiple input single output}
\newacronym{mpc}{MPC}{multi path component}
\newacronym{mrc}{MRC}{maximal ratio combining}
\newacronym{ni}{NI}{National Instruments}
\newacronym{nlos}{NLOS}{non-line of sight}
\newacronym{ntnu}{NTNU}{Norwegian University of Technology and Science}
\newacronym{ofdm}{OFDM}{orthogonal frequency division multiplexing}
\newacronym{pdf}{PDF}{probability density function}
\newacronym{reranp}{ReRaNP}{Reconfigurable Radio Network Platform}
\newacronym{rms}{rms}{root mean square}
\newacronym{rx}{RX}{receiver}
\newacronym{simo}{SIMO}{single input multiple output}
\newacronym{siso}{SISO}{single input single output}
\newacronym{snr}{SNR}{signal to noise ratio}
\newacronym{tdd}{TDD}{time division duplex}
\newacronym{tx}{TX}{transmitter}
\newacronym{ue}{UE}{user equipment}
\newacronym{ul}{UL}{uplink}
\newacronym{wsn}{WSN}{wireless sensor network}

\begin{document}
\bstctlcite{IEEEexample:BSTcontrol} 
\title{Measured Channel Hardening \\in an Indoor Multiband Scenario}
                                    
\author{
\IEEEauthorblockN{
Golsa~Ghiaasi\IEEEauthorrefmark{1}, %
Jens~Abraham, %
Egil~Eide and %
Torbj{\"o}rn~Ekman %
}
\IEEEauthorblockA{\\%
Department of Electronic Systems,
Norwegian University of Science and Technology,
Trondheim, Norway\\%
Email: \IEEEauthorrefmark{1}golsa.ghiaasi@ntnu.no
}}

\maketitle 
\widowpenalty10000 

\begin{abstract} 

A study of channel hardening in a large-scale antenna system has been carried out by means of indoor channel measurements over four frequency bands, namely \SI{1.472}{\giga\hertz}, \SI{2.6}{\giga\hertz}, \SI{3.82}{\giga\hertz} and \SI{4.16}{\giga\hertz}.
NTNU's \glsdesc{reranp} has been used to record the channel estimates for 40 single user \glsdesc{nlos} radio links to a 64 element wide-band antenna array. By examining the \glstext{rms} delay spread and the ratio of the normalized subcarrier to average \glstext{siso} link power of the radio channel received by a single user after combination, the hardening of this equivalent channel is analyzed for various numbers of nodes. The channel hardening merits show consistent behaviour throughout the frequency bands. By combining 16 antennas the \glstext{rms} delay spread of the equivalent channel is reduced from above \SI{100}{\nano\second} to below \SI{35}{\nano\second} with significantly reduced variation in the channel power. 
\end{abstract}

\section{Introduction}
In the view of the realization of \gls{tdd} massive \gls{mimo} testbeds at Lund University and University of Bristol \cite{malkowsky_worlds_2017}, the capabilities of such systems have been subject to investigative work, so as to do a real-life evaluation of the expected properties described by theoretical work such as \cite{rusek_scaling_2013, ngo_no_2017, bjornson_massive_2016, ngo_aspects_2014}. 
Most of the reported work has been focused on investigating measures of properties suitable for multi-user scenarios such as spectral efficiency, capacity and sum-rates  \cite{harris_throughput_2016,gao_massive_2015} as well as user separation and channel orthogonality through evaluation of singular value spread \cite{harris_performance_2017, flordelis_spatial_2015}.
The reported experimental results have been carried out at a single frequency band using antenna arrays composed of close and uniformly spaced antennas. Throughout the above-mentioned work, the concept of massive \gls{mimo} has been defined as a multi-user system with \gls{tdd} operation.    

In the topic of examining the channel properties over various frequency bands, the authors of \cite{li_measurement-based_2016} have characterized the measured channels for two indoor scenarios at three frequency bands in terms of path loss, delay spread and coherence bandwidth. The impact of the frequency on those parameters in \gls{siso} channels formed between \gls{tx} and \gls{rx} nodes have been discussed.

As an alternative approach to multi-user massive \gls{mimo}, the work in \cite{el-sallabi_experimental_2010} formulates the concept of an equivalent received channel at a single \gls{ue} by using  time-reversal filters along with the combination applied to  measured channels between the \gls{ue} and the \gls{bs}. The delay spread has been used as a measure for temporal focusing and the distribution of power around the strongest tap has been shown as the channel hardening measure. 
 
 Using the measured channels between a 128 antenna \gls{bs} and 36 \glspl{ue}, the authors in \cite{payami_delay_2013} have evaluated the \gls{rms} delay spread of the equivalent combined channel for three linear pre-coding schemes. 
 
In this paper, what we refer to the perceived channel at a single user in the \gls{dl} as an \textit{equivalent channel}. This is the channel constructed by pre-filtering of the signal followed by the combination at the antenna array in order to compensate for the channel, provided that reciprocity holds for the transmitter and receiver chains. 

We consider channel hardening as a property that causes the equivalent \gls{ctf} between the \gls{ue} and the \gls{bs} to become more deterministic with an increasing number of \gls{bs} antennas.
Harding of the channel can be exploited in \glspl{wsn} in order to ensure reliable communication between the \gls{bs} and the a number of low-complexity radios at the sensors.  
To achieve the reliability, it is required that the signal to each sensor propagates through a quasi-deterministic equivalent channel. At the same time, it is desirable to deploy radios at the sensor links with no need for complex equalization and estimation. 
In other words, it is advantageous to have simple \glspl{rx} by ensuring single tap realizations of the equivalent channel with a delay dispersion below the tap delay resolution, hence, eliminating the need for an equalizer at the sensor while exploiting channel hardening to reduce the variability of the signal strength of the equivalent channel and to ensure a strong signal at the sensor.
 
 We examine the channel hardening in the equivalent channel received at single \glspl{ue} from the \gls{bs} after matched-filter-weighted combination. We characterize the delay spread and the ratio of normalized subcarrier to average \gls{siso} power for different node combinations, namely 1, 4, 16 and 64 nodes, in order to determine the measure of flatness of the equivalent channel at the \gls{ue}. Knowing how many antennas are sufficient to achieve a certain level of channel hardening allows us to consider the remaining \gls{bs} antennas as contributors to achieve a multi-user system by using the remaining degrees of freedom to orthogonalize the equivalent channels.

First, we report on a channel measurement campaign which was carried out in an indoor area with industrial profile in a quasi-static scenario in 40 spatial sample points in order to characterize \gls{nlos} \gls{ul} channels to a 64 node array in 4 frequency bands: \SI{1.472}{\giga\hertz}, \SI{2.6}{\giga\hertz}, \SI{3.82}{\giga\hertz} and \SI{4.16}{\giga\hertz}. 
Second, the acquired \gls{ul} channel estimates are analyzed to formulate the equivalent channel perceived at single \glspl{ue}, that is, the measured \gls{siso} channels are matched filtered and combined in order to construct the observed channel at the \gls{ue}. Third, we examine the \gls{rms} delay spread and the ratio of the normalized subcarrier to average \gls{siso} power of the equivalent channel as our measures of channel hardening for 4 frequency bands while combining different number of nodes. 

This manuscript is organized as follows: in section \ref{sec:setup}, the channel measurement campaign is documented. A description of the measured scenario is presented in section \ref{sec:scenario}. The analysis of the measured data and formulation of channel hardening are described in section \ref{sec:channelhardening} with results of the comparison presented in section \ref{sec:measurement}. Lastly, the conclusions are presented in Section \ref{sec:Concolusion}.

\section{Measurement Description}
\label{sec:setup}
The investigation reported in this paper is based on a measurement campaign carried out at \gls{ntnu} in December 2017 using the \gls{reranp}.
The details of the \glspl{ue} and the \gls{bs} along with channel estimate acquisition in our settings are described in the following subsections: 

\subsection{Massive MIMO Prototype Hardware}
The massive \gls{mimo} testbed housed at \gls{ntnu} is composed of 4 units which each contain 32 radio chains by employing 16 \gls{ni} USRP-2943R devices.
These units controlled through a CPU which is running \gls{ni} LabVIEW Communications \gls{mimo} Application Framework  
implement a \gls{tdd} system with a \gls{lte}-like physical layer with \SI{20}{\mega\hertz} bandwidth of operation\cite{noauthor_introduction_2017}.
The key parameters of the system are summarized in Table \ref{tab:system_parameter}.
In this setting, 64 RF chains placed in 2 units (out of 4) were used.

\begin{table}[!t]
\centering
\caption{Massive \gls{mimo} Testbed Parameters}
\begin{tabular}{lr} 
\toprule
Parameter & Value  \\ 
\midrule
\# of \gls{bs} antennas & up to 128 (64 used)  \\ 
\# of \glspl{ue} & 8  \\ 
Center Frequency &  \SIrange{1.2}{6}{\giga\hertz} \\ 
Bandwidth of Operation & \SI{20}{\mega\hertz}  \\ 
Baseband Sampling Rate & \SI[per-mode=symbol]{30.72}{\mega\sample\per\second}\\
Subcarrier Spacing &  \SI{15}{\kilo\hertz}\\
\# of Subcarriers & 1200\\
FFT size & 2048\\
Frame Duration & \SI{10}{\milli\second}\\
Subframe Duration& \SI{1}{\milli\second}\\
Slot Duration  & \SI{0.5}{\milli\second}\\
\gls{tdd} periodicity & \SI{1}{\milli\second}\\
\bottomrule
\end{tabular}
\label{tab:system_parameter}
\end{table}
 
\subsection{Antennas}
\begin{figure*}
    \centering
    \includegraphics[trim={0cm 0cm 0cm 0cm},clip,width=.9\textwidth]{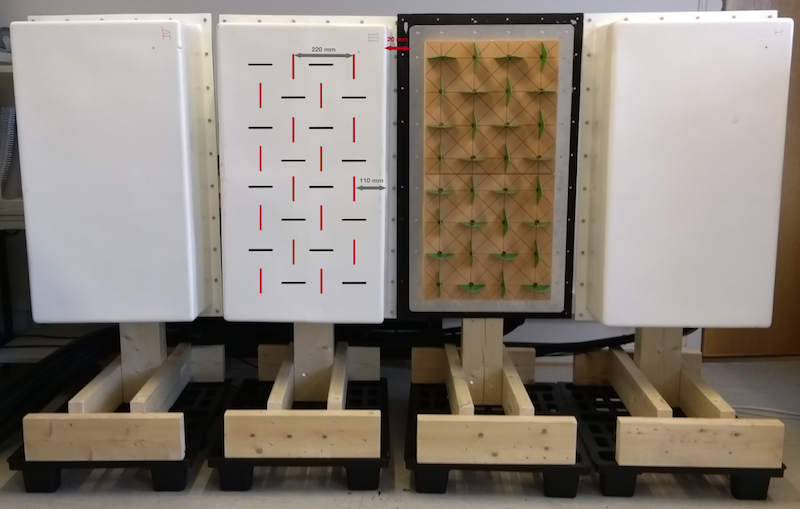}
    
    \caption{Frontal view of the subarray element configuration where only the vertically oriented elements were used.}
    \label{fig:array}
\end{figure*}

The \gls{bs} antenna array consists of wide-band \glspl{lpda} covering the frequency band between \SIrange{1.3}{6.0}{\giga\hertz}.
The linearly polarized \gls{lpda} element has been designed to provide a half power beamwidth of approximately \SI{110}{\degree} in the H-plane and \SI{70}{\degree} in the E-plane giving a directive gain of \SI{6}{\deci\bel i} when used as a single element.
The \glspl{ue} are equipped with one antenna element each, whilst the \gls{bs} is equipped with 4 subarrays each containing 32 elements in an equally spaced 4x8 rectangular configuration as illustrated in Fig. \ref{fig:array}.
The \glspl{lpda} are mounted with an element spacing of \SI{110}{\milli\meter} on a common ground plane.
As shown in Fig. \ref{fig:array} the antenna elements have interleaved polarization such that each element has a neighbor with orthogonal polarization.
This reduces the mutual coupling effects between the elements to a minimum, hence the effect on the input impedance and radiation pattern for the elements are minimized.
However, the introduction of the ground plane behind the \glspl{lpda} will give some variations in the directive gain during the measurements.
64 vertically polarized elements were terminated at the \gls{bs} whilst the other 64 horizontally polarized elements were left open.
The configuration is shown in detail in Fig. \ref{fig:array}.

\subsection{User Equipment Terminals}

The two radio chains of each \gls{ni} USRP-2953R unit \cite{noauthor_usrp-2953_nodate} along with \gls{ni} LabVIEW Communications MIMO Application Framework set up in Mobile configuration enable operation of two \glspl{ue}.
In our test we used 4 units resulting in a total of 8 \glspl{ue}.
The USRPs were arranged on a wheeled cart in a semi-circle configuration as shown in Fig. \ref{fig:array_ue} to facilitate only \gls{nlos} links in the scenario.

\subsection{Channel Estimate Acquisition}
The system has $M$ antennas on the \gls{bs} side, $K$ \glspl{ue} and uses \gls{ofdm} with 1200 usable subcarriers distributed over a \SI{20}{\mega\hertz} band. 
Each user $k$ transmits pilot symbols on a different subcarrier subset (covering 100 subcarriers) during the channel estimate acquisition to ensure orthogonality between them.
Collecting channel estimates from different physical locations in the same area allows to increase the number of measured realizations.

The complex \gls{ctf} coefficient $H_{mk}[l]$, with $k$, $l$ and $m$ as user index, subcarrier index (of the user subcarrier subset with size $L$) and \gls{bs} antenna index, respectively, is calculated by a least-squares estimate of the received signal with known pilot symbols.
To differentiate the true \gls{ctf} coefficient $H_{mk}[l]$ from the estimated \gls{ctf} coefficient a tilde is added on top of the variable: $\tilde{H}_{mk}[l]$.

\section{Measurement Scenario}
\label{sec:scenario}

\begin{figure*}[!t]
    \centering
    \includegraphics[trim={0cm 3cm 0cm 10cm},clip,width=.95\textwidth]{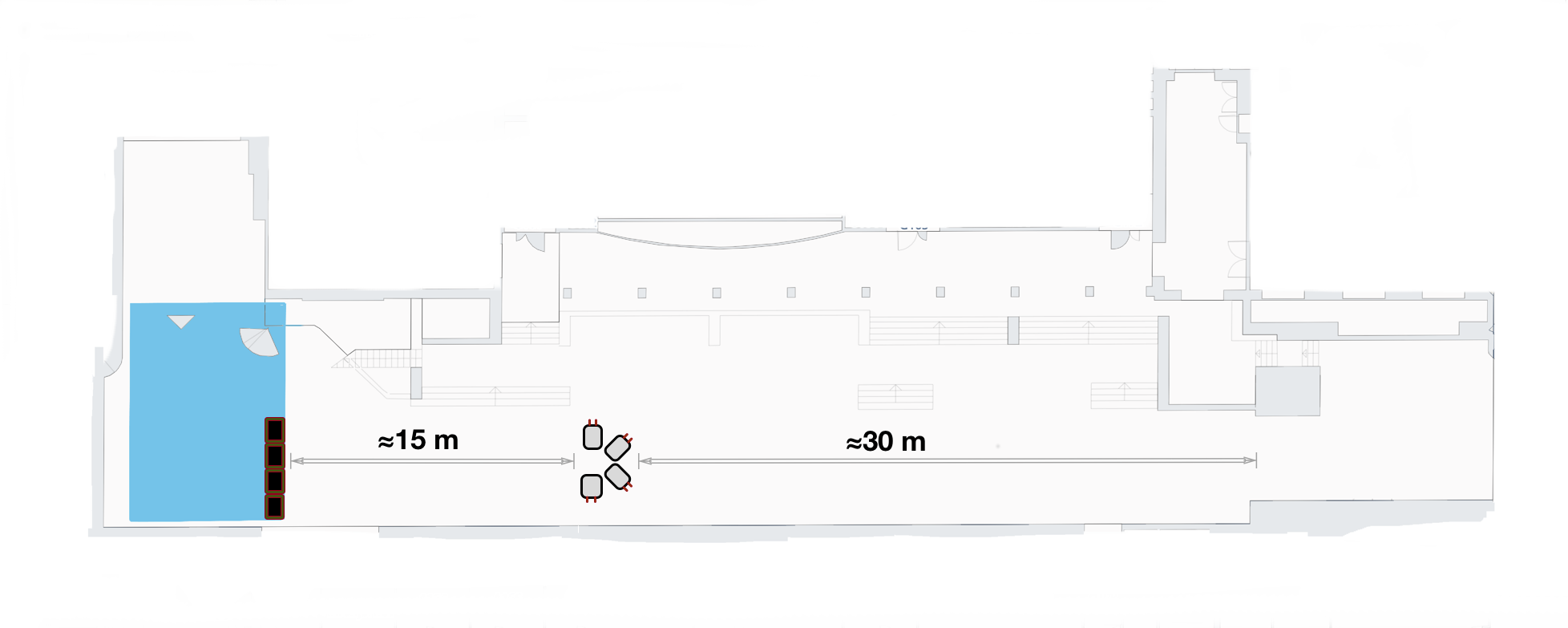}
    \caption{Top down sketch of the measurement scenario. The \glspl{ue} were placed at a distance of approximately \SI{15}{\meter} with antenna orientation away from the \gls{bs} subarrays.}
    \label{fig:floorplan}
\end{figure*}

\begin{figure}[!t]
    \begin{subfigure}[b]{.45\textwidth}
        \centering
        \includegraphics[trim={0 4cm 0 2cm},clip,width=.9\textwidth]{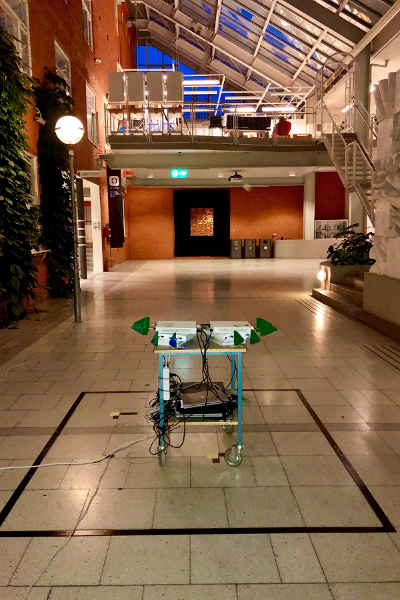}
        \caption{Positioning of the base station and UE cart}
        \label{fig:array_ue}
    \end{subfigure}
    \begin{subfigure}[b]{.45\textwidth}
        \centering
        \includegraphics[trim={0 0cm 0 3cm},clip,width=.9\textwidth]{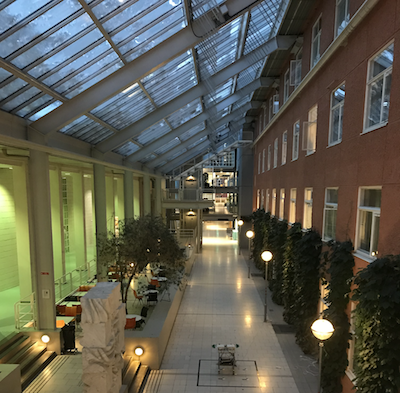}
        \caption{\glspl{ue} pointing towards the far end of the hall}
        \label{fig:ue_nlos}
    \end{subfigure}
    \caption{Indoor measurement campaign photos}
    \label{fig:test_setup}
\end{figure}

The measurement campaign was carried out in \gls{nlos} configuration in an indoor space with industrial profile.
As depicted in Fig. \ref{fig:floorplan} the BS was set up at the balcony at the end of the long hall while the cart containing the 8 \glspl{ue} was placed at around 15 meter distance from the \gls{bs}.
The antennas of the \glspl{ue} are positioned in a half-circle, directed away from the \gls{bs} to ensure \gls{nlos}.
The approximate distance between the \gls{ue} cart and the main reflectors at the end of the hall is around \SI{30}{\meter}.
As shown in the campaign photos in Fig. \ref{fig:test_setup}, there exist many reflecting surfaces.
Among others, concrete walls, window glasses, metal lamp posts and metal bars (at the end of the hallway) which form a rich scattering environment.
For each frequency band, the \gls{ue} cart was positioned along 5 pre-marked locations within \SI{1}{\meter} diameter to gain more sampling points of the environment.  

\section{Channel Hardening Metrics}
\label{sec:channelhardening}

On one hand, channel hardening can be seen as broadband flat fading of the equivalent \gls{ctf} in the frequency domain.
On the other hand, the equivalent \gls{cir} reveals \gls{ue} design requirements in the delay domain, since the \gls{rms} delay spread indicates the necessity of an equalizer.
The \gls{rx} could be simplified, if strong contributions of the perceived \gls{cir} are confined in a single delay tap.
Additionally, channel hardening would ensure that this tap power is stable and close to deterministic for most realizations.

Weighted sums of \gls{siso} \gls{ctf} coefficients with freely chosen weights $W_{mk}[l]$ are used to calculate equivalent \gls{ctf} coefficients $\hat{H}_{k}[l]$ (hat highlighting an equivalent channel variable)
\begin{equation}
    \hat{H}_{k}[l] = \sum_{m \in \hat{M}} W_{mk}[l] \tilde{H}_{mk}[l] \label{eqn:equiv_tfc}
\end{equation}
where $k$, $l$, $m$ represent the user, subcarrier and \gls{bs} antenna in the subset $\hat{M}$, respectively.
The latter allows the consideration of several equivalent channels to determine how the \gls{rms} delay spread behaves for an increasing number of \gls{bs} antennas.
We chose the \gls{mrc} approach (being optimal in a single user \gls{snr} sense) and used the complex conjugate of the estimated \gls{ctf} coefficients ($W_{mk}[l] = \tilde{H}_{mk}^\ast[l]$) as weights.
Finally, the \gls{cir} coefficients are calculated with an \gls{idft}
\begin{equation}
    \hat{h}_{k}[n] = \idft_l\left\{ \hat{H}_{k}[l]\right\} \label{eqn:equiv_irc}
\end{equation}
where $n$ denotes the delay bin, to represent an equivalent tapped delay line.

Moving the summation in Eqn. \eqref{eqn:equiv_tfc} outside of the \gls{idft} in Eqn. \eqref{eqn:equiv_irc} reveals that the equivalent \gls{cir} is the sum over Fourier transformed weighted \gls{siso} \glspl{ctf}.
Further, the equivalent \gls{cir} can be seen as sum over convolutions between the Fourier transformed weights and the \gls{siso} \glspl{cir}.
As a consequence, the Fourier transformed weights act as filtering in the delay domain.
The choice of \gls{mrc} weights forces this filter to be the time reversed and conjugated \gls{cir} of the estimated \gls{siso} channel.
Hence, the equivalent \gls{cir} becomes a sum over the auto-correlations of those channels.

Lastly, the \gls{rms} delay spread of an equivalent \gls{cir} is calculated by means of the normalized second-order central moment of $\lvert\hat{h}_k[n]\rvert^2$ for different subset choices of the available \gls{bs} antennas and form our first metric.

The second metric investigates the power variations between different subcarriers in the observed bandwidth.
We are normalizing the subcarrier power for a specific user $k$ with the number of receive antennas of a subset $\#\hat{M}$ to exclude the array gain.
Additionally, the resulting power is set in relationship to the average received \gls{siso} power of the specific user to have a normalized ratio and thus highlighting the variations only:
\begin{equation}
    \hat{R}_{k}[l] = \frac{
        \dfrac{1}{\#\hat{M}} \abs{\sum\limits_{m \in \hat{M}} W_{mk}[l] \tilde{H}_{mk}[l]}
    }{
        \dfrac{1}{ML} \sum\limits_{m=0}^{M-1} \sum\limits_{l=0}^{L-1} \abs{\tilde{H}_{mk}[l]}^2
    }. \label{eqn:subcarrier_power_metric}
\end{equation}

Statistics of $\hat{R}_{k}[l]$ allow to characterize the differences a narrow band \gls{rx} will see in the frequency range in terms of power levels with respect to the average received power.
Hence, it allows us to draw conclusions about the link budget in a direct way.

\section{Measurement Results}
\label{sec:measurement}
Measurements of $\tilde{H}_{mk}[l]$ were done for the aforementioned frequency bands in 5 physical locations with 8 \glspl{ue} probing the \gls{siso} channels.
The data of two receiver chains where the \gls{snr} indicated faulty equipment and failing links for \glspl{ue} with synchronization issues were excluded.

\begin{figure}[!t]
    \centering
    \includegraphics[width=0.48\textwidth]{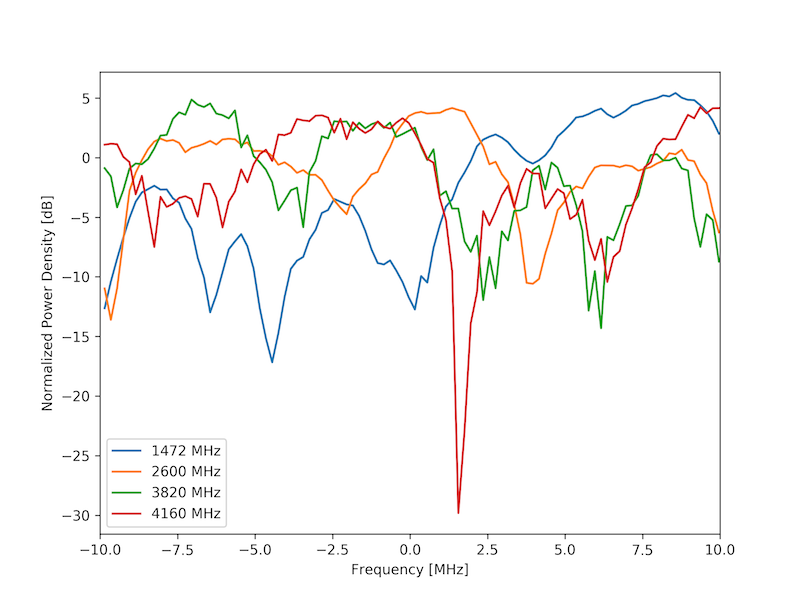}
    \caption{Selected transfer functions of a \gls{siso} uplink channel normalized to the average power.
    Fading of certain frequencies can be observed in each band.}
    \label{fig:norm_tfs}
\end{figure}

To get an overview over the measured radio environment, a single \gls{ctf} per frequency band is shown in Fig. \ref{fig:norm_tfs}.
They were normalized to their average power level to highlight the fluctuations over the frequency range.
Significant fading below \SI{-10}{\deci\bel} can be found, as expected for a rich scattering environment, in each of them.

\begin{figure}[!t]
    \centering
    \includegraphics[width=0.48\textwidth]{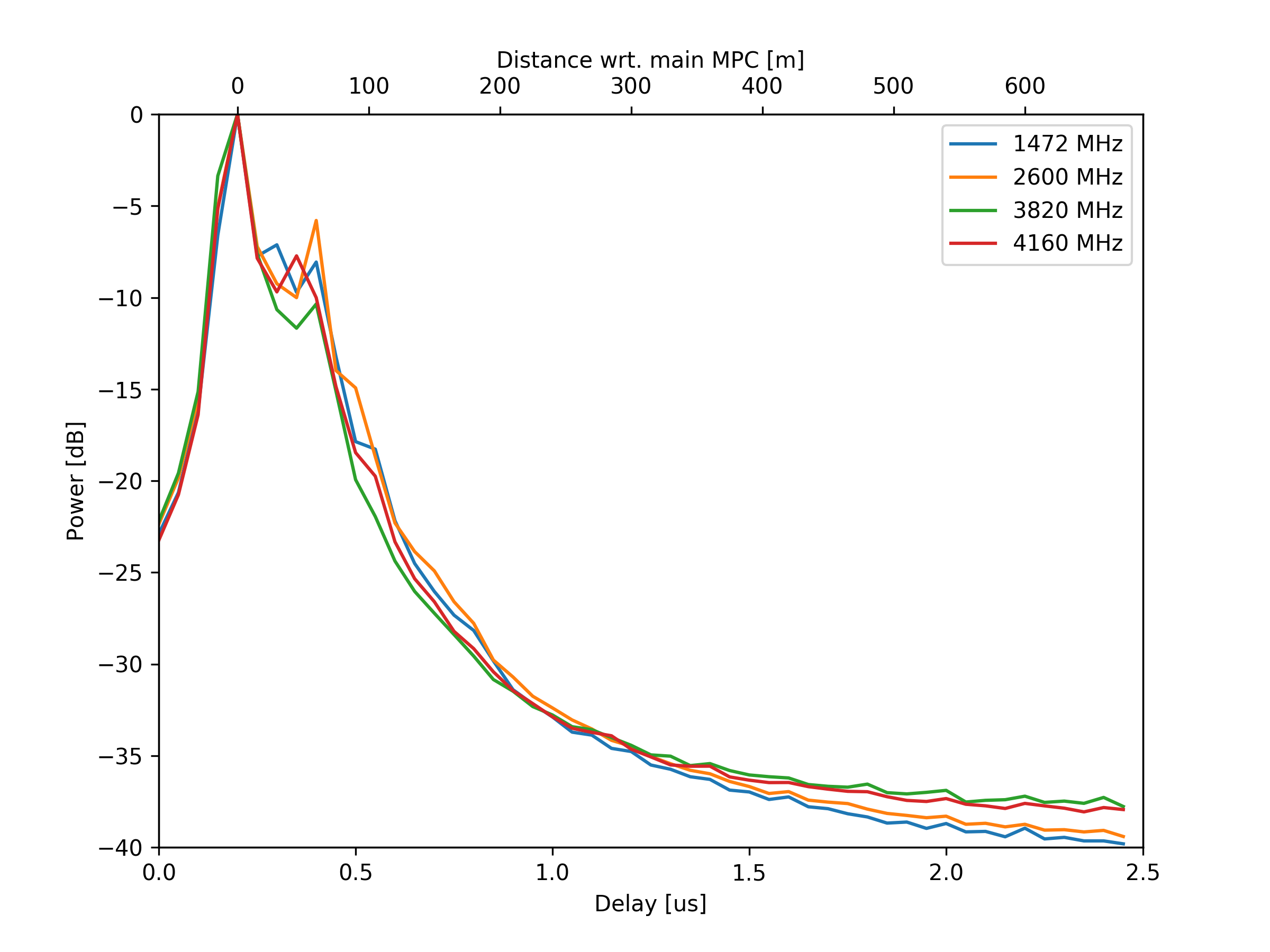}
    \caption{Normalized power delay profiles for the four frequency bands. The power delay profile was averaged over all realizations between the different users and base station antennas and most \gls{mpc} contributions are confined to a delay window of \SI{1}{\micro\second}.}
    \label{fig:norm_pdps}
\end{figure}

Fig. \ref{fig:norm_pdps} represents normalized power delay profiles for all measured \gls{siso} channels by averaging over the realizations for all \glspl{ue} and \gls{bs} antennas.
At least two \glspl{mpc} are clearly resolved for the observation bandwidth of the user.
Nonetheless, most power is confined in a delay window of \SI{1}{\micro\second}.
Considering that minimal power is captured outside of that window gives rise to suppress everything outside of a \SI{+-1}{\micro\second} window for the equivalent channel.
Artifacts of the \gls{idft} at the delay border and noise contributions are therefore reduced.

\begin{figure*}[!t]
    \centering
    \includegraphics[width=\textwidth]{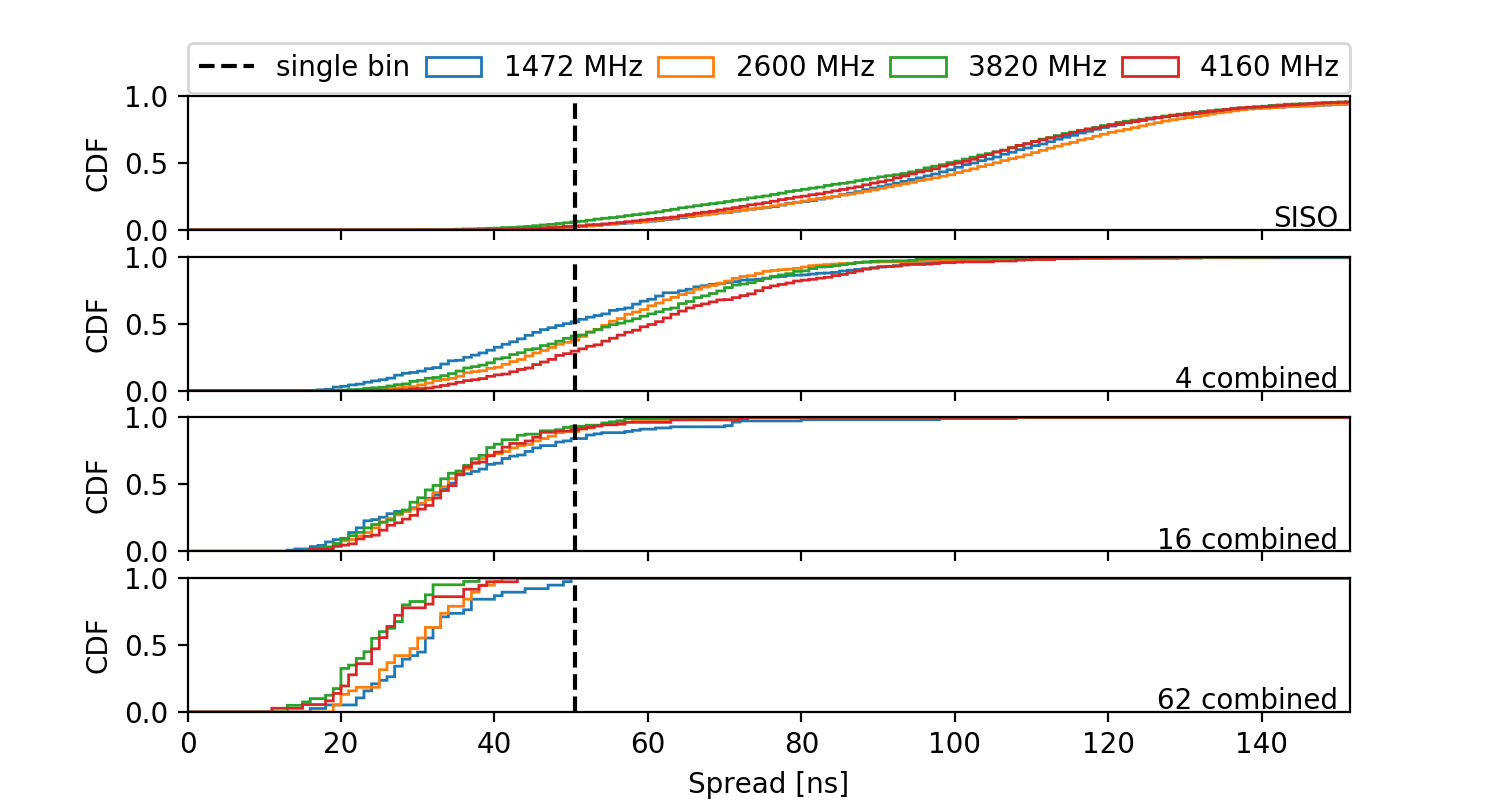}
    \caption{Cumulative distribution functions of the \gls{rms} delay spread for different numbers of combined \gls{siso} channels.
    With an increasing number of combined channels decreases the spread towards a single bin width.
    Using just 16 antennas for combining, more then \SI{80}{\percent} of the equivalent channel realizations have a \gls{rms} delay spread below the single time bin.}
    \label{fig:combined_spreads}
\end{figure*}

Fig. \ref{fig:combined_spreads} shows the histogram approximated \gls{cdf} of the \gls{rms} delay spread.
All considered bands for the \gls{siso} case as well as combinations of 4, 16 and 62 antennas are shown.
The combinations for the smaller numbers were drawn from similar subsets of consecutive close antenna elements in the subarrays.
The two excluded \gls{rx} reduce the number of realizations slightly but the overall behaviour for an increasing number of combined antennas is still captured.

\begin{figure}[!t]
    \centering
    \includegraphics[width=0.48\textwidth]{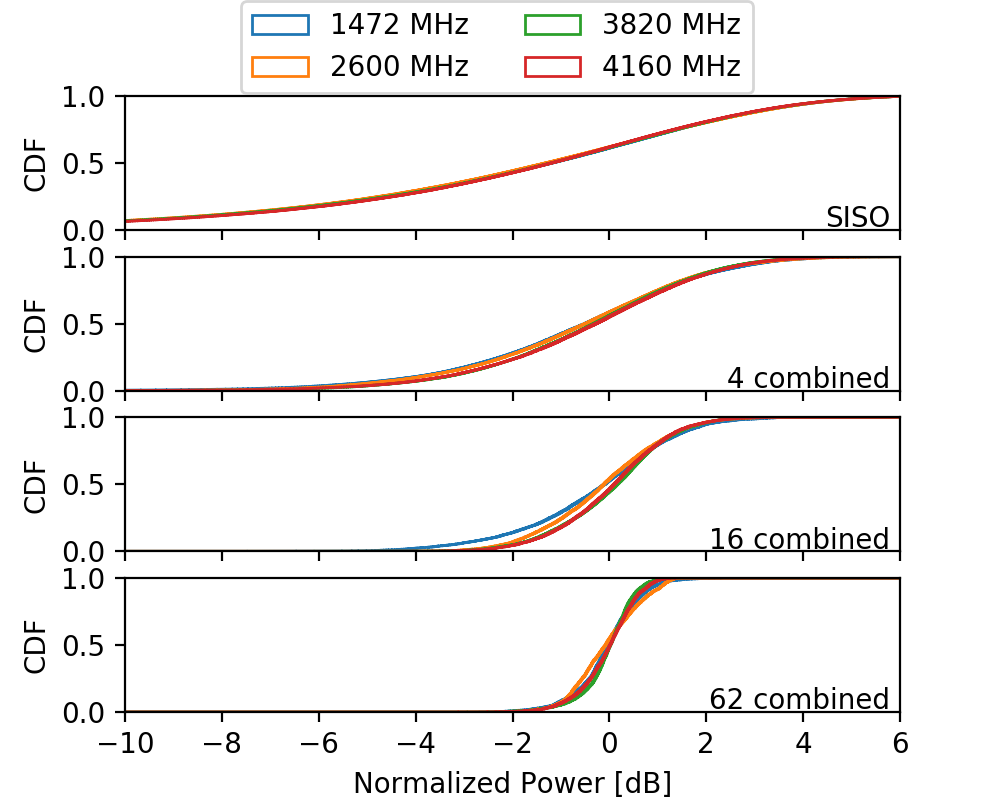}
    \caption{Cumulative distribution functions of the ratio of the normalized subcarrier to average \gls{siso} power in logarithmic units.
    The variations in power levels between different subcarriers decrease as the number of combined receivers is increased.
    The combination of 16 receiver chains restricts the fading level to above \SI{-3.5}{\deci\bel} and the usage of all available receiver chains shows no fading below \SI{-1}{\deci\bel} for all subcarriers in \SI{95}{\percent} of the realizations.}
    \label{fig:combined_power}
\end{figure}

Similarly, Fig. \ref{fig:combined_power} shows the behaviour of the ratio of the normalized subcarrier to average \gls{siso} power for the respective \gls{ue}.
This allows us to use all observations as realizations for the histogram approximated \gls{cdf} of all $\hat{R}_{k}[l]$.
The drawn combinations are the same as in the \gls{rms} delay spread analysis.

\section{Conclusion}
\label{sec:Concolusion}

We have investigated a quasi-static indoor radio environment over four frequency bands in the range from \SIrange{1.4}{4.2}{\giga\hertz}.
The measurements contain 40 single user realizations against a 64 element antenna array \gls{bs} to elucidate channel hardening in both the delay and the frequency domain.
The \gls{rms} delay spread and the normalized subcarrier to average \gls{siso} power ratio of the equivalent radio channel were described as physically motivated figures of merit for the \gls{wsn} case.

Both merits are showing similar behaviour over all measured frequency bands.
Considering the inter-element spacing of the antenna arrays (about $\SI{155}{\milli\meter}$ between used \glspl{lpda}) being already larger then half the wavelength at the lowest frequency band, gives rise to the assumption that the correlation between the elements at the \gls{bs} is sufficiently small for all measured cases.
Hence, the spatial sampling of the incoming waves is de-correlated enough to show similar properties in the equivalent channel for all bands.

Analysis of the equivalent channel delay spread shows that 16 antennas configured in a single subarray are a good compromise to allow for a single tap \gls{rx} on the \gls{ue} side for the considered bandwidth whilst the remainder of the large \gls{bs} array can be considered advantageous for a multi-user massive \gls{mimo} system.
Similar conclusions can be drawn for the reliability of links over the equivalent channel since their observed fading dips approach a lower bound of about \SI{-4}{\deci\bel} for 16 antennas.
The latter allows to relax the fading margin of a link budget for a \gls{wsn} and can be used to reduce power requirements at the sensor side without sacrificing reliability constraints.

To summarize, channel hardening has been found to be a substantial property of \gls{simo}/\gls{miso} equivalent channels showing a remarkable impact by combining only 16 antennas.
Further, there was no significant frequency dependency observed over the considered bands with the \gls{reranp} testbed.
Future work will investigate the multi-user capabilities for the reported scenario.

\bibliographystyle{IEEEtran}

\bibliography{IEEEabrv,2018_pimrc.bib}

\begin{thebibliography}{10}
\providecommand{\url}[1]{#1}
\csname url@samestyle\endcsname
\providecommand{\newblock}{\relax}
\providecommand{\bibinfo}[2]{#2}
\providecommand{\BIBentrySTDinterwordspacing}{\spaceskip=0pt\relax}
\providecommand{\BIBentryALTinterwordstretchfactor}{4}
\providecommand{\BIBentryALTinterwordspacing}{\spaceskip=\fontdimen2\font plus
\BIBentryALTinterwordstretchfactor\fontdimen3\font minus
  \fontdimen4\font\relax}
\providecommand{\BIBforeignlanguage}[2]{{%
\expandafter\ifx\csname l@#1\endcsname\relax
\typeout{** WARNING: IEEEtran.bst: No hyphenation pattern has been}%
\typeout{** loaded for the language `#1'. Using the pattern for}%
\typeout{** the default language instead.}%
\else
\language=\csname l@#1\endcsname
\fi
#2}}
\providecommand{\BIBdecl}{\relax}
\BIBdecl

\bibitem{malkowsky_worlds_2017}
S.~Malkowsky, J.~Vieira, L.~Liu, P.~Harris, K.~Nieman, N.~Kundargi, I.~C. Wong,
  F.~Tufvesson, V.~Owall, and O.~Edfors, ``The {World}’s {First}
  {Real}-{Time} {Testbed} for {Massive} {MIMO}: {Design}, {Implementation}, and
  {Validation},'' \emph{{IEEE} Access}, vol.~5, pp. 9073--9088, 2017.

\bibitem{rusek_scaling_2013}
F.~Rusek, D.~Persson, {Buon Kiong Lau}, E.~G. Larsson, T.~L. Marzetta, and
  F.~Tufvesson, ``Scaling {Up} {MIMO}: {Opportunities} and {Challenges} with
  {Very} {Large} {Arrays},'' \emph{{IEEE} Signal Process. Mag.}, vol.~30,
  no.~1, pp. 40--60, Jan. 2013.

\bibitem{ngo_no_2017}
H.~Q. Ngo and E.~G. Larsson, ``No {Downlink} {Pilots} {Are} {Needed} in {TDD}
  {Massive} {MIMO},'' \emph{{IEEE} Trans. Wireless Commun.}, vol.~16, no.~5,
  pp. 2921--2935, May 2017.

\bibitem{bjornson_massive_2016}
E.~Björnson, E.~G. Larsson, and T.~L. Marzetta, ``Massive {MIMO}: ten myths
  and one critical question,'' \emph{{IEEE} Commun. Mag.}, vol.~54, no.~2, pp.
  114--123, Feb. 2016.

\bibitem{ngo_aspects_2014}
H.~Q. Ngo, E.~G. Larsson, and T.~L. Marzetta, ``Aspects of favorable
  propagation in {Massive} {MIMO},'' in \emph{Proceedings of the 22nd
  {European} {Signal} {Processing} {Conference} ({EUSIPCO}), 2014}, Lisbon,
  Portugal, Sep. 2014.

\bibitem{harris_throughput_2016}
P.~Harris, S.~Zhang, M.~Beach, E.~Mellios, A.~Nix, S.~Armour, A.~Doufexi,
  K.~Nieman, and N.~Kundargi, ``{LOS} {Throughput} {Measurements} in
  {Real}-{Time} with a 128-{Antenna} {Massive} {MIMO} {Testbed},'' in
  \emph{2016 IEEE {Global} {Communications} {Conference} ({GLOBECOM})}.\hskip
  1em plus 0.5em minus 0.4em\relax IEEE, Dec. 2016, pp. 1--7.

\bibitem{gao_massive_2015}
X.~Gao, O.~Edfors, F.~Rusek, and F.~Tufvesson, ``Massive {MIMO} {Performance}
  {Evaluation} {Based} on {Measured} {Propagation} {Data},'' \emph{{IEEE}
  Trans. Wireless Commun.}, vol.~14, no.~7, pp. 3899--3911, Jul. 2015.

\bibitem{harris_performance_2017}
P.~Harris, S.~Malkowsky, J.~Vieira, E.~Bengtsson, F.~Tufvesson, W.~B. Hasan,
  L.~Liu, M.~Beach, S.~Armour, and O.~Edfors, ``Performance {Characterization}
  of a {Real}-{Time} {Massive} {MIMO} {System} {With} {LOS} {Mobile}
  {Channels},'' \emph{{IEEE} J. Sel. Areas Commun.}, vol.~35, no.~6, pp.
  1244--1253, Jun. 2017.

\bibitem{flordelis_spatial_2015}
J.~Flordelis, X.~Gao, G.~Dahman, F.~Rusek, O.~Edfors, and F.~Tufvesson,
  ``Spatial separation of closely-spaced users in measured massive multi-user
  {MIMO} channels,'' in \emph{2015 IEEE International Conference on
  Communications (ICC)}.\hskip 1em plus 0.5em minus 0.4em\relax IEEE, Jun.
  2015, pp. 1441--1446.

\bibitem{li_measurement-based_2016}
J.~Li, B.~Ai, R.~He, K.~Guan, Q.~Wang, D.~Fei, Z.~Zhong, Z.~Zhao, D.~Miao, and
  H.~Guan, ``Measurement-{Based} {Characterizations} of {Indoor} {Massive}
  {MIMO} {Channels} at 2 {GHz}, 4 {GHz}, and 6 {GHz} {Frequency} {Bands},'' in
  \emph{2016 IEEE 83rd Vehicular Technology Conference (VTC Spring)}.\hskip 1em
  plus 0.5em minus 0.4em\relax IEEE, May 2016, pp. 1--5.

\bibitem{el-sallabi_experimental_2010}
H.~El-Sallabi, P.~Kyritsi, A.~Paulraj, and G.~Papanicolaou, ``Experimental
  {Investigation} on {Time} {Reversal} {Precoding} for {Space}–{Time}
  {Focusing} in {Wireless} {Communications},'' \emph{{IEEE} Trans. Instrum.
  Meas.}, vol.~59, no.~6, pp. 1537--1543, Jun. 2010.

\bibitem{payami_delay_2013}
S.~Payami and F.~Tufvesson, ``Delay spread properties in a measured massive
  {MIMO} system at 2.6 {GHz},'' in \emph{2013 IEEE 24th Annual International
  Symposium on Personal, Indoor, and Mobile Radio Communications
  (PIMRC)}.\hskip 1em plus 0.5em minus 0.4em\relax IEEE, Sep. 2013, pp. 53--57.

\bibitem{noauthor_introduction_2017}
\BIBentryALTinterwordspacing
``Introduction to the {NI} {MIMO} {Prototyping} {System} {Hardware},'' Dec.
  2017. [Online]. Available: \url{http://www.ni.com/white-paper/53197/en/}
\BIBentrySTDinterwordspacing

\bibitem{noauthor_usrp-2953_nodate}
\BIBentryALTinterwordspacing
``{USRP}-2953 {Specifications}.'' [Online]. Available:
  \url{http://www.ni.com/pdf/manuals/374197d.pdf}
\BIBentrySTDinterwordspacing

\end{thebibliography}

\end{document}